\documentclass[12pt]{article}
\usepackage{amsfonts,amsmath,amssymb,epsf}
\usepackage{graphicx,color}
\newcommand{\al}{\alpha'}

\newcommand{\be}{\begin{equation}}
\newcommand{\ba}{\begin{eqnarray}}
\newcommand{\ea}{\end{eqnarray}}
\newcommand{\ee}{\end{equation}}

\newcommand{\f}{\frac}
\newcommand{\s}{\sqrt}

\newcommand{\ti}{\tilde}

\newcommand{\ddd}{\cdot\cdot\cdot}

\newcommand{\tabtopsp}[1]{\vbox{\vbox to#1{}\vbox to1mm{}}}

\begin{document}

\begin{titlepage}
\thispagestyle{empty}

hep-th/0702194

KUNS-2063

\begin{flushright}
\end{flushright}

\bigskip

\begin{center}
\noindent{\Large \textbf{Free Yang-Mills\ \  vs.
Toric Sasaki-Einstein}}\\
\vspace{2cm} \noindent{Tatsuma
Nishioka\footnote{e-mail:nishioka@gauge.scphys.kyoto-u.ac.jp} and
Tadashi
Takayanagi\footnote{e-mail:takayana@gauge.scphys.kyoto-u.ac.jp}}\\
\vspace{1cm}

 {\it  Department of Physics, Kyoto University, Kyoto 606-8502, Japan }

\vskip 2em
\end{center}

\begin{abstract}

It has been known that the Bekenstein-Hawking entropy of the black
hole in $AdS_5\times S^5$ agrees with the free $\mathcal{N}=4$ super
Yang-Mills entropy up to the famous factor $\f{4}{3}$. This factor
can be interpreted as the ratio of the entropy of the free
Yang-Mills to the entropy of the strongly coupled Yang-Mills. In
this paper we compute an analogous factor for infinitely many
$\mathcal{N}=1$ SCFTs which are dual to toric Sasaki-Einstein
manifolds. We observed that this ratio always takes values within a
narrow range around $\f{4}{3}$. We also present explicit values of
volumes and central charges for new classes of toric Sasaki-Einstein
manifolds.

\end{abstract}

\end{titlepage}

\newpage

\section{Introduction}
\setcounter{equation}{0}

The AdS/CFT correspondence \cite{Maldacena} has been playing a
crucial role to explore the non-perturbative aspects of gravity and
gauge theories for ten years. Even though there are many important
examples of AdS/CFT, a general condition that a given CFT should
have its AdS dual has not been known completely until now\footnote{
A well-known necessary condition is that the two central charges $a$
and $c$ in 4D CFT should be equal $a=c$ in order to have its AdS
dual as far as we consider the AdS backgrounds known so far
\cite{HeSk}.}. One simple strategy to understand this issue better
is to study many examples of AdS/CFT systematically and see if there
are any common properties for CFTs which have their AdS duals.

Fortunately, infinitely many $\mathcal{N}=1$ superconformal gauge
theories have recently been known to have their AdS duals in terms
of the five dimensional Sasaki-Einstein manifolds $X_5$
\cite{GMSW,BFHMS,CLPP}, generalizing the celebrated example
$T^{1,1}$ \cite{KlWi}. Therefore it is very interesting to find any
common properties among them. A basic and physically important
quantity will be the degrees of freedom of a given CFT. We can
estimate this by computing its thermodynamical entropy at finite
temperature\footnote{Another way to measure the degrees of freedom
will be to count the BPS states of a given SCFT. This has been
discussed in \cite{GG}, recently.}.

We can easily compute the entropy of a super Yang-Mills theory in
its strong coupling limit as the entropy of the dual black hole
\cite{GuKlPe,Wi}. On the other hand, it is very difficult to
calculate the entropy directly in the strongly coupled Yang-Mills
theory. Instead we assume a free Yang-Mills approximation of SCFTs.
This crude approximation works better than we naively expect, owing
to the super conformal symmetry. Indeed it has been known that this
approximation deviates from the dual gravity result only by the
factor $\f{4}{3}$ in the $\mathcal{N}=4$ super Yang-Mills
\cite{GuKlPe}. Since this semi-quantitative agreement is a
remarkable property, it is intriguing to see if a similar agreement
is true for other SCFTs which have their AdS duals. Furthermore, it
has been pointed out that even the Hagedorn transition in the dual
string theory on $AdS_5$ can also be captured from the free
Yang-Mills theory \cite{Hage}.

The aim of this paper is to investigate this ratio
$\f{S_{free}}{S_{strong}}$ of the entropy of various $\mathcal{N}=1$
quiver gauge theories with all interactions turned off, to the
entropy in the strongly coupled $\mathcal{N}=1$ SCFTs realized as IR
fixed points of the interacting quiver theories. We can physically
interpret this ratio as a measure of the strength of interactions in
a given CFT. We can compute the black hole entropy, which is
inversely proportional to the volume of Sasaki-Einstein manifolds,
by employing the Z-minimization method \cite{MaSpYa} dual to the
a-maximization \cite{IW,BZ,Tac,BGIW}. Therefore we can obtain this
ratio only from the information of the toric data for any toric
Sasaki-Einstein manifolds. After we search large families of
infinitely many toric diagrams, we find that the ratio is always in
a narrow range $\f{8}{9}\leq \f{3}{4}\cdot
\f{S_{free}}{S_{strong}}\lesssim 1.2$. The minimum value
$\f{S_{free}}{S_{strong}}=\f{32}{27}$ is realized when $X_5$ is
equal to $T^{1,1}$ or its orbifolds.

This paper is organized as follows. In section 2 we give the
expression of the ratio $\f{S_{free}}{S_{strong}}$ in terms of the
volume of Sasaki-Einstein manifolds and the number of fields in the
dual gauge theory. In section 3 we calculate this ratio explicitly
for various examples. In section 4 we compute an analogous ratio in
$\mathcal{N}=1$ SQCD. In section 5, we summarize our results and
also discuss other interesting quantities.

\section{Entropy from Black Hole and Free Yang-Mills}
\setcounter{equation}{0}

Consider a background $AdS_5\times X_5$ in type IIB supergravity,
where $X_5$ is a five dimensional Sasaki-Einstein manifold. This
theory is dual to a four dimensional $\mathcal{N}=1$ SCFT
\cite{KlWi}. Such a theory is explicitly described by a $SU(N)$
quiver gauge theory \cite{KlWi,BFHMS}. A systematic construction of
such gauge theories is recently found by using brane-tiling method
\cite{FHKVW,FHMSVW} (for recent progresses see e.g.
\cite{HV,FHKV,Im,IIKY}).

 The AdS radius is found to be \be
R=\left(\f{4\pi^4g_s\al^2N}{\mbox{Vol}(X_5)}\right)^{\f14}, \ee
where Vol$(X_5)$ is the volume of $X_5$ normalized such that
Vol$(S^5)=\pi^3$ \cite{HeSk,Gu}. The central charge $a$ of the SCFT
\cite{Cardy} is related to the volume via \be
a=\f{N^2}{4}\cdot\f{\pi^3}{\mbox{Vol}(X_5)}. \ee

The thermodynamical entropy $S$ in the strong coupling limit of the
SCFT can be found from the entropy of AdS-Schwarzschild black
hole\footnote{In this paper we consider the AdS black holes in the
Poincare coordinate.} ($T$ is the temperature) \be
S_{strong}=\f{\mbox{Horizon Area}}{4G^{(5)}_{N}} =\f{\pi^5
N^2}{2\mbox{Vol}(X_5)}VT^3 =2\pi^2a VT^3, \ee where $G^{(5)}_N$ is
the 5D Newton constant and it is related to the 10D Newton constant
$G^{(10)}_N=8\pi^6\al^4 g_s^2$ via the dimensional reduction
$G^{(10)}_N=R^5 \mbox{Vol}(X_5)\cdot G^{(5)}_N$. The entropy is
proportional to the central charge and this is consistent with the
expectation that $a$ is related to the degrees of freedom.

However, we would like to notice that the entropy is not always
proportional to $a$ for all coupling regions. For example, in the
free super Yang-Mills theory, the entropy is proportional to the
number of bosons $N_B$ as the contribution of a (free) gauge field
$A^\mu$ is the same as a (free) complex scalar field $\Phi$ (we
count each of these as a unit $N_B=1$.). Since the central charge
$a$ of $A^\mu$ is different from that of $\Phi$, the free Yang-Mills
entropy is not proportional to $a$.

Therefore it is interesting to consider the ratio
$\f{S_{free}}{S_{strong}}$ of the free Yang-Mills entropy to the
strongly coupled Yang-Mills entropy. This ratio measures how the
degrees of freedom changes when we turn on the interactions of the
quiver gauge theories. It is well-known that
$\f{S_{free}}{S_{strong}}$ becomes $\f{4}{3}$ for the
$\mathcal{N}=4$ super Yang-Mills theory \cite{GuKlPe}. In general
N=1 SCFTs, we have to worry about the ambiguity of the field content
of the free field approximation due to the Seiberg
duality\footnote{We are very grateful to
 Yuji Tachikawa for pointing this issue to us in detail.}. Even
though the entropy in the free Yang-Mills depends on the frame of
Seiberg duality or equally on the choice of toric phases of $X_5$
\cite{BHK}, we will find that this ambiguity changes the entropy
only slightly in explicit examples. Thus this does not spoil our
semi-quantitative argument in this paper. We will proceed the
arguments by choosing a standard toric phase.

In the end, this ratio can be computed as follows via AdS/CFT \be
\f{S_{free}}{S_{strong}}=\f{4}{3}\cdot\f{N_B}{4N^2}\cdot
\f{\mbox{Vol}(X_5)}{\pi^3}=\f{4}{3}\cdot \f{N_B}{16a}\equiv
\f{4}{3}f, \label{ratios} \ee where we defined the ratio $f$
normalized such that $f=1$ for the $\mathcal{N}=4$ super Yang-Mills
theory. We will present results below in term of this ratio $f$. A
larger value of $f$ means that the degrees of freedom is more
reduced in the strongly coupled regime compared with the free
Yang-Mills. As we will see later, $f$ takes values of order $1$
(i.e. $\sim (1\pm 0.2)$) in all examples we studied. In the orbifold
theories, we always find $f=1$. Moreover, the value of $f$ remains
the same after we take a $Z_n$ orbifold of any Sasaki-Einstein
manifold $X_5$. Notice that the confirmation that the ratio is
always of order one is already a non-trivial check of the AdS/CFT
duality for infinitely many SCFTs.

We can represent other physical quantities in terms of $f$. The
analogous ratio $\f{E_{free}}{E_{strong}}$ of the Casimir energy in
a super Yang-Mills compactified on a thermal circle \cite{HoMy} is
the same as before $\f{E_{free}}{E_{strong}}=\f{4}{3}f$. Also the
ratio of entanglement entropy \cite{RyTaL} becomes
$\f{S^{free}_A}{S^{strong}_A}=\f{2}{3}f$ \cite{NiTa} when we define
the subsystem $A$ by dividing the boundary into two half planes.

We would like to stress again that the ratio $f$ is essentially (the
inverse of) the central charge $a$ divided by the number of free
fields $N_B$. The central charge itself increases (linearly) as the
size of toric diagram grows\footnote{This can also be seen in
explicit examples. It is clear from the result (\ref{volypq}) for
$Y^{p,q}$ that the corresponding central charge behaves as $a\propto
p$ in the limit $p\to \infty$ with $p/q$ kept finite.} \cite{Ka}.
Since the number of fields $N_{B}$ also scales linearly as the area
of the diagram becomes infinitely large, the ratio $f$ stays finite.

\section{Entropy from Toric Sasaki-Einstein Manifolds}
\setcounter{equation}{0}

A classification of the toric Sasaki-Einstein manifolds can be
obtained by using the toric diagrams \cite{MaSpYa,FuOn} which
describe corresponding Calabi-Yau cones. Though originally the toric
diagrams are three dimensional for the cones over $X_5$, we can
project them onto a two dimensional plane owing to the Calabi-Yau
condition. Thus we can write the coordinates of vertices in the
toric diagrams with $n$-vertices (i.e. $n$-polygon) as
$(1,p_i,q_i)\in {\bf Z}^3$ $(i=1,2,\ddd,n)$. We can compute several
physical quantities of a quiver gauge theory from the toric diagram
of its dual Sasaki-Einstein manifold.

The number of vector multiplets and chiral multiplets in a standard
choice of the toric phase are given by \cite{BFHMS}
\be
N_{gauge}=2\cdot(\mbox{Area of the toric diagram}),\ \ \ \ \
N_{matter}=\sum_{1\leq i<j\leq n}|p_iq_j-p_jq_i|. \label{nmat}
 \ee The above formula shows that $N_{gauge}$ is the same as
 the Euler number of the toric manifold. This is because the Euler number
counts the number of independent (fractional) D-branes. On the other
hand, $N_{matter}$ is found by considering intersections of 3-cycles
in the mirror Calabi-Yau cone in \cite{BFHMS}.

 Using this, the number of total bosons is found to be
 $N_B=N_{gauge}+N_{matter}$. Notice that in the above formula
 (\ref{nmat}) of the number of matter fields $N_{matter}$, we are
 choosing a particular toric phase. We will be able to employ
 other equivalent
 descriptions of the quiver gauge theories by applying the Seiberg's
 duality where the number of matter fields
 $N_{matter}$ takes different values \cite{BHK}. As we will see below in
 the explicit example of $Y^{p,q}$, this ambiguity changes
 the value of the ratio $f$ only slightly, though we cannot give a
 complete general discussion on this issue since its systematic treatment has not
 been developed well at present. Thus we will compute the ratio $f$
 by choosing a particular choice of the toric phase by using the formula
 (\ref{nmat}) in most of examples in this paper.
 We argue that the ambiguity of the toric phases does
 not spoil our semi-quantitative arguments in this paper because it
 does not affects the ratio $f$ substantially as mentioned. Also notice that
 in spite of this subtle ambiguity of the physical definition of the ratio
 $f$, this quantity is completely well-defined after we plug
 (\ref{nmat}) into the formula. This means that this value is
 mathematically exactly well-defined as we can find a unique value when a
 toric diagram is given. Thus this quantity is also very interesting from
 this mathematical viewpoint.

Quite recently, the existence and uniqueness of the Sasaki-Einstein
metric have been proved in \cite{FuOn} if a given toric diagram
satisfies a simple condition (such a toric diagram is called good)
\footnote{This condition essentially requires that $p_i-p_{i+1}$ and
$q_i-q_{i+1}$ are coprime for all $i$ (refer to the second paper of
\cite{FuOn}).}. First we will study all toric diagrams described by
four vertices and then we examine some particular classes with five
or more vertices. To summarize the results obtained in this section,
we will draw up the Table 1 and Table 2 at the end of this paper.

\subsection{Toric diagrams with Four Vertices}

\subsubsection{$Y^{p,q}$}

\begin{figure}[htbp]
  \begin{center}
    \includegraphics[scale=0.5]{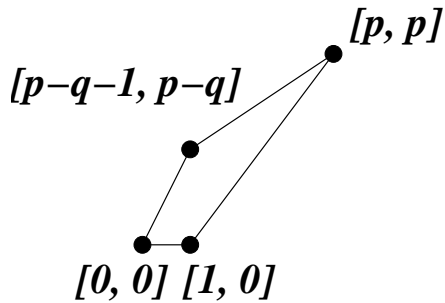}
  \end{center}

  \caption{Toric diagram of $Y^{p,q}$} \label{fig:torYpq}
\end{figure}

As a first example, we consider the familiar example $Y^{p,q}$
\cite{GMSW} whose toric diagram is given by Fig.\ref{fig:torYpq}
($p$ and $q$ are integers such that $p\geq q\geq 0$). The volume of
$Y^{p,q}$ is known to be
\be \f{\text{Vol}(Y^{p,q})}{\pi^3} = \f{q^2(2p + \s{4p^2-3q^2})}
{3p^2(3q^2 - 2p^2 + p\s{4p^2-3q^2})}. \label{volypq} \ee

\begin{figure}[htbp]
  \begin{center}
    \includegraphics[scale=1]{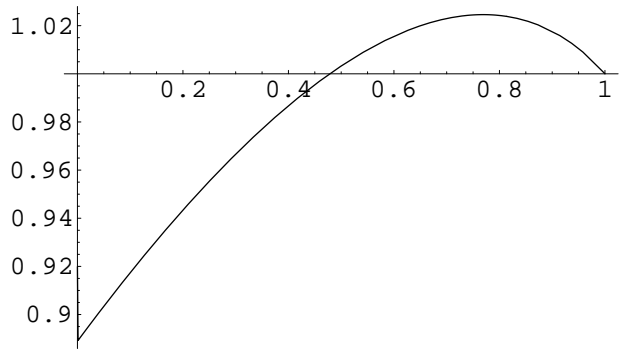}
  \end{center}

  \caption{The values of $f$ for $Y^{p,q}$}\label{fig:fYpq}
\end{figure}

The number of bosons which appear in the dual field theory
\cite{BFHMS} is given by $\f{N_B}{N^2} = 2(3p+q)$, where we have
employed the explicit values (\ref{nmat}) in the standard choice of
toric phase. As a result, the ratio $f$ is expressed as follows \be
f(x) = \f{x^2(3+x)(2+\s{4-3x^2})}{6(3x^2-2+\s{4-3x^2})}, \qquad x
\equiv \f{q}{p}. \ee This is plotted in Fig.\ref{fig:fYpq} and the
function $f(x)$ $(0\leq x\leq 1)$ takes the values within the range
\be \f{8}{9} \le f(x) \le 1.02459. \label{ypsr} \ee Notice that at
$x=0$, where $X_5$ becomes the orbifold $T^{1,1}/Z_p$,
 the function $f(x)$ takes its minimum value $f=\f{8}{9}$. The
 maximum value in (\ref{ypsr}) is attained when $x=0.76929$.
On the other hand, at $x=1$, where $X_5$ is the orbifold
$S^5/Z_{2p}$, the function $f(x)$ takes $f=1$.

Now, let us ask how the ratio $f$ depends on the choice of toric
phases. All toric phases in $Y^{p,q}$ have been obtained in
\cite{BHK}. The number of matter fields takes the following range
\be 6p+2q\leq \f{N_B}{N^2}\leq 8p, \ee if we search all toric
phases. The lowest value is the one (\ref{nmat}) we employed in the
above and this leads to the result (\ref{ypsr}).
 On the other hand, if we compute the ratio $f$ by using the
 highest possible value $\f{N_B}{N^2}=8p$, then the ratio takes the
 following range
\be 1<f_{high}<\f{32}{27}. \ee This deviates from (\ref{ypsr}) less
than twenty percent. Thus we expect that our particular choice
(\ref{nmat}) of the toric phase does not spoil our semi-quantitative
arguments in this paper.

\subsubsection{$L^{p,q,r}$}

\begin{figure}[htbp]
  \begin{center}
    \includegraphics[scale=0.5]{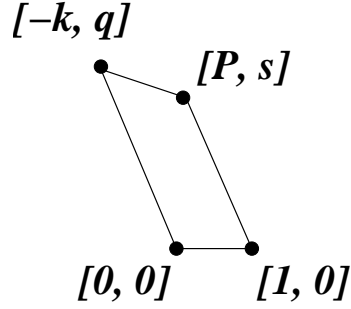}
  \end{center}

  \caption{Toric diagram of $L^{p,q,r}$. The integers $p,q,r,s$ and
  $P$ are taken such that $p+q = r+s,\quad ks+qP=r,\quad k > 0$ and
     $0\le p \le r \le s \le q$.  } \label{fig:torLpqr}
\end{figure}

We move onto the toric Sasaki-Einstein manifolds $L^{p,q,r}$. This
class includes all examples whose toric diagrams consist of four
vertices as shown in Fig.\ref{fig:torLpqr}. Define $x \equiv
\f{p}{q},~y \equiv \f{r}{q}$ and then they should satisfy $0\le x
\le y \le 1, ~y\le \f{x+1}{2}$. The relation between $Y^{p,q}$ and $
L^{p,q,r}$ is given
 by $Y^{p,q} = L^{p-q, p+q, p}$ and thus $Y^{p,q}$ is on the line $y= \f{x+1}{2}$.
The volume of $L^{p,q,r}$ is found to be as follows \cite{CLPP} \be
\text{Vol}(L^{p,q,r}) = \pi^3\cdot\f{(p+q)^3 W}{8pqrs}, \ee where
$W$ is the solution to the quartic equation \ba
&&(1-F^2)(1-G^2)h_-^4 + 2h_-^2\left[ 2(2-h_+)^2 -3h_-^2 \right] W
\nonumber\\
&&\quad + \left[ 8h_+(2-h_+)^2 - h_-^2(30+9h_+)\right] W^2 +
8(2-9h_+)W^3 - 27W^4 =0. \ea Here we defined $F=\f{1-x}{1+x},\quad
G=\f{2y-x-1}{1+x}, \quad h_\pm = F^2 \pm G^2. $ There are four
solutions to this equation and only one of them is a positive real
number. We use this solution for computing the volume of
$L^{p,q,r}$.

The number of bosons in the dual gauge theory \cite{PLpqr,FHMSVW} is
given by $\f{N_B}{N^2} = 2(p+2q)$. In the end, the ratio $f(x,y)$ is
expressed as follows \be f(x,y) =
\f{(1+x)^3(2+x)W(x,y)}{xy(1+x-y)},\qquad ( x \le y,~ y \le
\f{x+1}{2} ). \ee The numerical analysis shows that the range of
$f(x,y)$ is the same as the one for $Y^{p,q}$
 \be \f{8}{9} \le
f(x,y) \le 1.02459. \ee

\subsection{Toric diagrams with Five Vertices}

\subsubsection{$X^{p,q}$}
\begin{figure}[htbp]
  \begin{center}
    \includegraphics[scale=0.5]{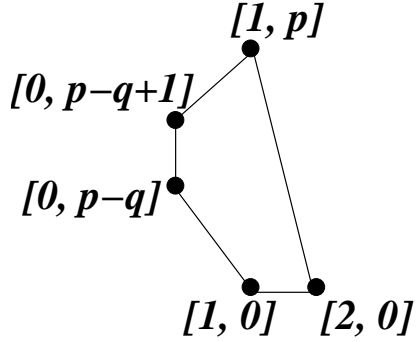}
  \end{center}

  \caption{Toric diagram of $X^{p,q}$}\label{fig:torXpq}
\end{figure}

$X^{p,q}$ is given by a specific toric diagram with five vertices
(Fig.\ref{fig:torXpq}) \cite{HKW}. The number of bosons are
$\f{N_B}{N^2}=6p+2q+2$. We can compute the volume of $X^{p,q}$ using
the Z-minimization procedure \cite{MaSpYa}, but its analytic
expression is difficult to find. Therefore we performed a numerical
analysis and obtained the following range \be \f{8}{9} \le f \le
1.03729. \ee The maximum is realized when $(p,q)=(4,3)$.

\subsubsection{Symmetric Pentagon}

\begin{figure}[htbp]
  \begin{center}
    \includegraphics[scale=0.5]{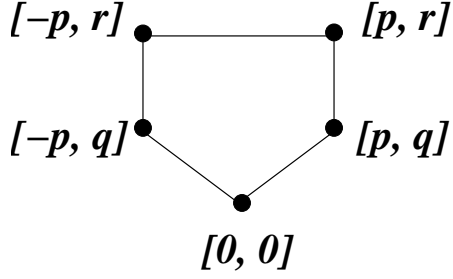}
  \end{center}

  \caption{Toric diagram of symmetric pentagon} \label{fig:torddP2}
\end{figure}

To find many more examples, we consider the five vertex toric
diagram defined by Fig.\ref{fig:torddP2}. It has the reflection
symmetry along the vertical axis\footnote{This symmetry allowed us
to set the Reeb vector of the form $(3,0,z)$, which largely reduces
the amount of the computations of Z-minimization.}. Note also that
this toric diagram can be regarded as a deformation of that of the
second delPezzo surface $dP_2$. The numbers of fields are
$\f{N_B}{N^2}=12pr-4pq$. Notice that since integer points are
included on the edges of toric diagram, the resulting manifold also
includes orbifold singularities\footnote{This means that these toric
diagrams do not generically satisfy the `goodness' condition in
\cite{FuOn}. However, the corresponding manifolds and their dual
gauge theories are physically sensible as the orbifolds
$T^{1,1}/Z_{p}$ are. Also if we would like to keep the diagrams
good, we can consider limits of good toric diagrams. If we take the
coordinates $(p_i,q_i)$ very large, we can approach the symmetric
pentagon as much as we want because the function $f$ is invariant
under the total scaling of the diagram.}\label{dptwo}.

 We can perform the Z-minimization analytically in this case and the
 Reeb vector is found as
 $\left(3,0,\f{3(4qr-r^2-\s{8qr^3+r^4}\ )}{4(q-r)}\right)$.
 Then $f$ is given
 in terms of $y\equiv \f{q}{r}$ $(0\leq y\leq 1)$ by
\be f(y) =
\f{32(3-y)(1-y)^3(1+\s{8y+1})}{27(3-\s{8y+1})(1+\s{8y+1}-4y)^2}. \ee

This expression shows us the function $f$ takes the range
 \be \f{8}{9} \le f(y) \le
1.03172. \ee Finally, we would like to summarize the central charge
and the R-charges of baryons in this example as follows \ba && a =
\f{27p(3r^2 -
  \s{8qr^3+r^4})(r^2-4qr+\s{8qr^3+r^4})^2}{128(r-q)^3(r^2+\s{8qr^3+r^4})}N^2,\nonumber\\
&&R_1 = R_4 =
\f{(1+\s{8y+1}-4y)(3-\s{8y+1})}{4(1-y)^2(1+\s{8y+1})}N,\nonumber\\
&&R_2 = R_3 =\f{(1+\s{8y+1}-4y)^2}{4(1-y)^2(1+\s{8y+1})}N,
 \quad R_5 = \f{2y(3-\s{8y+1})}{(1-x)(1+\s{8y+1})}N.
\ea

\subsection{Toric diagrams with Six Vertices}

\subsubsection{$Z^{p,q}$}
\begin{figure}[htbp]
  \begin{center}
    \includegraphics[scale=0.5]{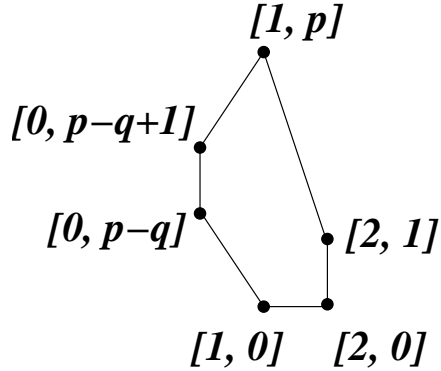}
  \end{center}

  \caption{Toric diagram of $Z^{p,q}$} \label{fig:torZpq}
\end{figure}

The Sasaki-Einstein manifold $Z^{p,q}$ is defined by the toric
diagram with six vertices Fig.\ref{fig:torZpq} \cite{OoYa}. The
number of bosons is $\f{N_{B}}{N^2}=6p+2q+4$.

The volume of $Z^{p,q}$ is given as follows \cite{OoYa} \be
\text{Vol}(Z^{p,q}) =
\pi^3\cdot\f{9p^3-9p^2q+6pqy_2-2y_2^2}{3y_2^2(3p-y_2)^2}. \ee Here
$y_2$ is the solution to the cubic equation $2y^3 - 9pqy^2 -
9p^2(2p-3q)y + 27p^3(p-q)=0$ which lives in the region $0 < y_2 <
3p$. We find the function $f$ takes following values \be \f{8}{9}
\le f \le 1.05007. \ee The maximum value is again taken for
$Z^{4,3}$.

\subsubsection{Symmetric Hexagon}
\begin{figure}[htbp]
  \begin{center}
    \includegraphics[scale=0.5]{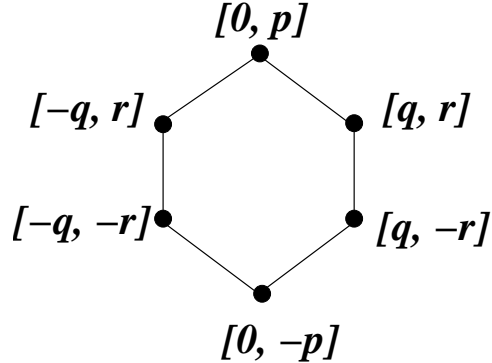}
  \end{center}

  \caption{Toric diagram of symmetric hexagon}  \label{fig:torhex}
\end{figure}

As another example of toric diagrams with six vertices\footnote{
Notice that as in the symmetric pentagon, the corresponding
manifolds will have orbifold singularities and the same footnote as
in section \ref{dptwo} applies.}, we consider the ones with the
reflection symmetries along the horizontal and vertical directions
described by Fig.\ref{fig:torhex}. This symmetry allows us to set
the Reeb vector equal to $(3,0,0)$. The number of bosons is
$\f{N_{B}}{N^2}=12q(p+r)$.
In this case, the function $f$ does not depend on $q$ as in the case
of the symmetric pentagon. Define $x \equiv \f{r}{p}$ $(0\leq x\leq
1)$ and then the analytic expression of the function $f(x)$ is
simply given by \be f(x) = \f{4(2-x)(1+x)}{9}. \ee Thus we obtain
\be \f{8}{9} \le f(x) \le 1. \ee The maximum value is taken when
$x=1/2$, which corresponds to the third delPezzo surface $dP_3$. The
minimum value is realized when $x=0,1$, i.e. orbifolds of $T^{1,1}$.

Note that the central charge and R-charges of baryons are given by
the following expressions \ba
&&a = \f{27p^2q}{16(2p-r)}N^2\nonumber\\
&&R_1 = R_2 = R_4 = R_5 = \f{1}{2(2-x)}N, \quad R_3 = R_6 = \f{1}{2-x}N.
\ea

\subsection{Toric Diagram with Infinitely Many Vertices}

It is very important to find how much the upper bound of the
function $f$ increases as we raise the number $n$ of vertices in
toric diagrams. As the easiest example for general $n$ we
concentrate on the specific example of the most symmetric toric
diagram (i.e. the regular polygons) whose vertices are given by
$\rho\left( \cos\left( \f{2\pi}{n}i\right), \sin\left(
    \f{2\pi}{n}i\right)\right),~(i=1,2,\dots,n)$. The value of $f$
    does not depend on $\rho$ because
    the function $f$ is invariant under the total rescaling of toric
    diagram.  Therefore, we can realize this
    diagram as a limit of very large toric
    diagrams taking $\rho\to\infty$. Then, because of the
$Z_2 \times Z_2$ symmetry, it is clear that the Reeb vector is given
by $(3, 0, 0)$. Thus it is direct to compute its volume and the
function $f$. The numbers of fields in the dual gauge theory are
$\f{N_{gauge}}{N^2}=n\sin\left( \f{2\pi}{n}\right)$ and
$\f{N_{matter}}{N^2}=2n\sin\left( \f{2\pi}{n}\right)$ for $n$ even,
$\f{N_{matter}}{N^2}=2n\sin\left( \f{\pi}{n}\right) \left( 1 +
\cos\left(
    \f{\pi}{n}\right)\right)$ for $n$ odd (we set $\rho=1$).
The fuction $f$ becomes \ba &&n\text{ : even} \qquad f(n) =
\f{n^2}{9}\sin^2\left(\f{\pi}{n}\right) \to
\f{\pi^2}{9}\quad (n \to \infty)\\
&&n\text{ : odd} \qquad~ f(n) = \f{n^2}{27}
\f{\sin^2\left(\f{\pi}{n}\right)}{\cos\left(\f{\pi}{n}\right)}
\left( 1 + 2\cos\left(\f{\pi}{n}\right)\right) \to \f{\pi^2}{9}\quad
(n \to \infty). \ea Thus it does not become so large even if we
increase the number of vertices. Its range is \be \f{8}{9} \le f(n)
< \f{\pi^2}{9}=1.09662 \ee The minimum value is attained when $n=4$,
i.e. $T^{1,1}$.

We summarized all of previous examples in this section in the Table
1 and 2 at the end of this paper. We also added a non-toric example
$dP_4$ and compared it with the toric counterpart $PdP_4$, whose
central charges were computed in \cite{BuFoZa}. We also examined a
new example whose toric diagram is described by a symmetric octagon.

\begin{table}[b]
  \begin{center}
    \begin{tabular}{|c||c|c|c|}
      \hline\tabtopsp{3mm}%
      SE manifolds & $f_{max}$ & $N_{gauge}/N^2$ & $N_{matter}/N^2$
      \\[.5mm]
      \hline\tabtopsp{5mm}%
      $Y^{p,q}$ & 1.02459 & $2p$ & $4p+2q$  \\[1.5mm]
      \hline\tabtopsp{5mm}%
      $L^{p,q,r}$ & 1.02459 & $p+q$  & $p+3q$ \\[1.5mm]
      \hline\tabtopsp{5mm}%
      $X^{p,q}$ & 1.03729 & $2p+1$ & $4p+2q+1$ \\[1.5mm]
      \hline\tabtopsp{5mm}%
      $Z^{p,q}$ & 1.05007  & $2p+2$ & $4p+2q+2$ \\[1.5mm]
      \hline\tabtopsp{5mm}%
      \begin{minipage}{10mm}
        \begin{center}
          \unitlength=1mm
          \includegraphics[scale=0.15]{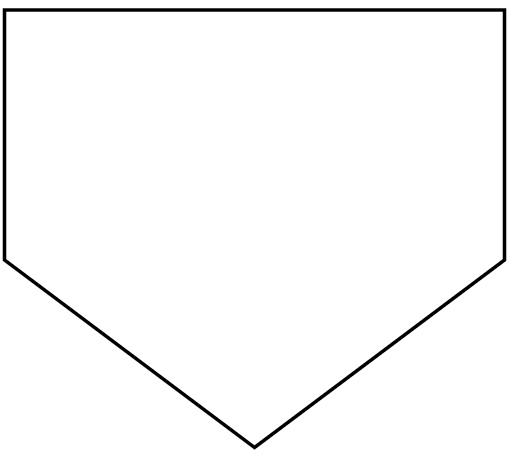}
        \end{center}
      \end{minipage}
      \ \  (symmetric pentagon)
      & 1.03172 & $2p(2r-q)$  & $2p(4r-q)$   \\ [2.5mm]
      \hline\tabtopsp{5mm}%
      \begin{minipage}{10mm}
        \begin{center}
          \unitlength=1mm
          \includegraphics[scale=0.15]{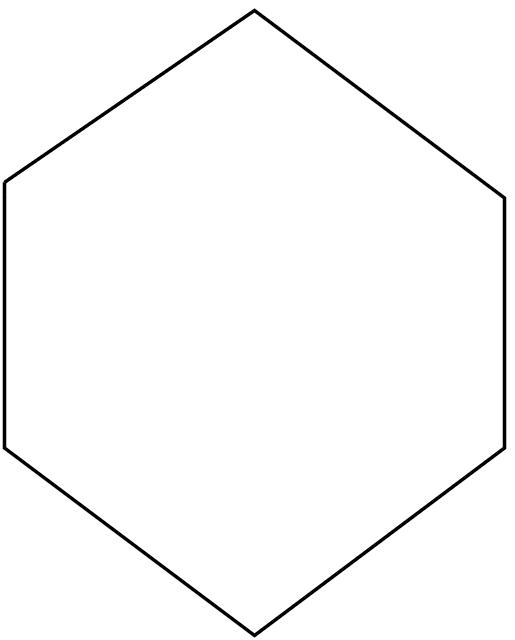}
        \end{center}
      \end{minipage}
      \ \ (symmetric hexagon)
      & 1  &  $4q(p+r)$  &  $8q(p+r)$   \\ [2.5mm]
      \hline\tabtopsp{6mm}%
      Regular polygon
      & 1.09662  &  $n\sin\left( \f{2\pi}{n} \right) $
      &  \parbox{5.7cm}{$2n\sin\left(\f{2\pi}{n}\right)$ ($n$:even)\\
        $2n\sin\left(\f{\pi}{n}\right)\left( 1+
          \cos\left(\f{\pi}{n}\right)
        \right)$ ($n$:odd)} \\ [3.5mm]
      \hline
    \end{tabular}
    \caption{Table of the values of $f$ for various Sasaki-Einstein
    manifolds considered in section 3.
      We gave the maximum value $f_{max}$ of $f$ as
      the minimum value is always $\f{8}{9}$.}
    \label{tablese}
    \vspace{15mm}
  \begin{tabular}{|c||c|c|c|c|}
      \hline\tabtopsp{3mm}%
      SE manifolds  & $f$ & Vol$(X_5)/\pi^3$ & $N_{gauge}/N^2$  & $N_{matter}/N^2$
      \\[.5mm]
      \hline\tabtopsp{5mm}%
      $dP_0=\mathbb{P}^2$\ ($S^5$)   & 1  &  $\f{1}{3}$        & $3$  & $9$  \\[1.5mm]
      \hline\tabtopsp{5mm}%
      $dP_1$\ ($Y^{2,1}$) & $\f{322+91\s{13}}{648}=1.00325$ &
       $\f{46+13\s{13}}{324}$   & $4$ & $10$ \\[1.5mm]
      \hline\tabtopsp{5mm}%
      $dP_2$\ ($X^{2,1}$)  & $\f{118+22\s{33}}{243}=1.00568$
      & $\f{59+11\s{33}}{486}$     &  $5$  & $11$  \\[1.5mm]
      \hline\tabtopsp{5mm}%
      $dP_3$\ ($Z^{2,1}$)  & 1 &   $\f{2}{9}$       & $6$    &  $12$ \\[1.5mm]
      \hline\tabtopsp{5mm}%
      SP\ ($X^{1,1}$) & $\f{5\s{3}}{9}=0.96225$ &  $\f{2\s{3}}{9}$   &   $3$  & $7$ \\[1.5mm]
      \hline\tabtopsp{5mm}%
      $PdP_4$ (toric)  & 0.96964  &    0.17630          &   7  & 15  \\[1.5mm]
      \hline\tabtopsp{5mm}%
      $dP_4$ (non-toric) & $\f{55}{54}=1.01852$ &  $\f{5}{27}$    & 7  & 15 \\[1.5mm]
      \hline\tabtopsp{6.5mm}%
      \begin{minipage}{10mm}
        \begin{center}
          \unitlength=1mm
          \includegraphics[scale=0.1]{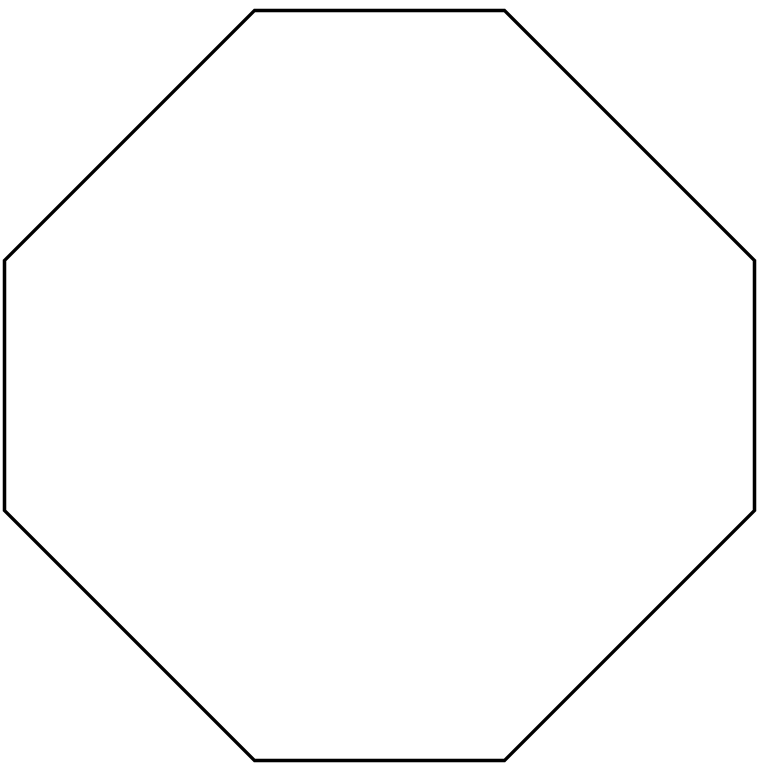}
        \end{center}
      \end{minipage}
  (symmetric octagon) & $\f{28}{27}=1.03704$ &  $\f{8}{81}$   & 14 & 28 \\[2.5mm]
      \hline
 \end{tabular}
    \caption{Table of the values of $f$ and the volumes
    for various Sasaki-Einstein
    manifolds defined by the specified four dimensional surfaces in
   the corresponding Calabi-Yau cones.
      The final example is defined by the octagonal toric diagram
       whose vertices are given by $(1,0)$,$(2,0)$,$(3,1)$,$(3,2)$,
       $(2,3)$,$(1,3)$,$(0,2)$ and $(0,1)$ after projected
        to the two dimensional plane $R^2$.}
    \label{tablese2}
  \end{center}
\end{table}

\section{Comparison with $\mathcal{N}=1$ SQCD}

The function $f$ can be calculated only from the gauge theoretic
data, i.e. the central charge and the number of bosons as is clear
from (\ref{ratios}). Therefore it will be useful to compare our
previous analysis for the $\mathcal{N}=1$ quiver gauge theories
which have AdS duals, with the one for the $\mathcal{N}=1$ SQCDs
whose AdS duals have not been known. Since we are interested in
SCFTs, we concentrate on the conformal window $\f{3}{2}N_c \le N_f
\le 3N_c$ of the SQCD with the $SU(N_c)$ gauge group and $SU(N_f)$
flavor group. It is well known that at the IR fixed point $2N_fN_c$
quarks $\psi_q, \psi_{\ti q}$ and $N_c^2-1$ gauginos $\lambda$ have
R-charge $R(\psi_q)=R(\psi_{\ti q})=-\f{N_c}{N_f}$ and $R(\lambda) =
1$ respectively. Thus the central charge \cite{AFGJ,AEFJ} in the IR
fixed point is given by \ba a^{IR} =
\f{3}{32}(3\mbox{Tr}R^3-\mbox{Tr}R)= \f{3}{16}\left(2N_c^2 - 1 -
\f{3N_c^4}{N_f^2}\right). \ea

Then we find the function $f$ is given by \be f = \f{2N_fN_c +
N_c^2-1}{6N_c^2 - 3 -\f{9N_c^4}{N_f^2}}. \ee In the planar limit
$N_c>>1$ setting $x\equiv \f{N_f}{N_c}$ finite, the function $f(x)$
is expressed by \be f(x) = \f{x^2(1+2x)}{6x^2-9}. \ee In the
conformal window $\f{3}{2} \le x \le 3$, this takes \be 1.30248 \le
f(x) \le 2. \ee The maximum and minimum values are taken for $x=3/2$
and $x=2.27163$ respectively.

\section{Conclusions and Discussions}
\setcounter{equation}{0}

In this paper we have studied the ratio
$\f{S_{free}}{S_{strong}}=\f{4}{3}f$ of the free Yang-Mills entropy
to the entropy in the strongly coupled $\mathcal{N}=1$ SCFTs via the
AdS/CFT correspondence. We mainly considered $\mathcal{N}=1$ SCFTs
dual to  toric Sasaki-Einstein manifolds $X_5$. Since they are
classified by toric diagrams, we could compute the ratio rather
systematically for infinitely many examples, though we could not
exhaust all toric Sasaki-Einstein manifolds. We checked that in all
examples the ratio takes the finite values within a rather narrow
range (for a standard choice of the toric phase) \be \f{8}{9}\leq f
\leq f_{max}. \label{boundf} \ee For example, the values of $f$ for
$Y^{p,q}$ and $L^{p,q,r}$ are included in the range $\f{8}{9}\leq
f\leq 1.02459$. We can  think remarkable even the fact that the
ratio takes a finite value of order one. Even though the central
charge $a$ is often used to characterize a given SCFT, it can take
any arbitrary large values.

{}From the infinitely many examples which we explicitly examined in
this paper, we can find the maximal value $f_{ourmax}=1.09662$. This
maximal value, however, seems to increase if we include other
examples of toric diagrams which we did not consider in this paper.
We would like to conjecture that the true upper bound $f_{max}$ is
only slightly larger than $f_{ourmax}$, say $f_{max}\sim 1.2$. We
gave an evidence for this behavior by presenting an explicit
analysis when the toric diagram is the regular $n$-polygon. On the
other hand, the lowest bound $f=\f{8}{9}$ is realized when $X_5$ is
(an orbifold of) $T^{1,1}$. Also notice that since
$\f{S_{free}}{S_{strong}}=\f{4}{3}f$ is always greater than $1$, the
degrees of freedom in a strongly coupled Yang-Mills is smaller than
those in the free Yang-Mills. This is natural since the interactions
generally give masses to the off-diagonal elements of massless
fields.

In most of the computations, we assumed a particular choice
(\ref{nmat}) of toric phase. The ratio $f$ takes slightly different
values when we employ different phases. We checked that this
ambiguity does not spoil our semi-quantitative argument in the
explicit example of $Y^{p,q}$. To understand this issue in detail we
need to develop a systematical way to analyze various toric phases
and we left it as a important future problem.

Our result strongly suggests that the entropy ratios for all such
$\mathcal{N}=1$ SCFTs deviate from the ones for the $\mathcal{N}=4$
super Yang-Mills (i.e. $f_{N=4}=1$) only by a small amount ($\sim
\pm 20\%$). This means that the degrees of freedom of the strongly
coupled SCFTs are not so different from the ones obtained in their
free Yang-Mills counterparts. This may be a bit surprising since we
know that such a $\mathcal{N}=1$ SCFT is realized as a non-trivial
IR fixed point of an interacting $\mathcal{N}=1$ quiver gauge theory
\cite{KlWi}. Indeed, we have seen that in the conformal window of
$\mathcal{N}=1$ SQCD, $f$ takes slightly larger values than the
range (\ref{boundf}).

 We would like to mention a possibility that our result
(\ref{boundf}) may be a special property which is common to all
$\mathcal{N}=1$ SCFTs with AdS duals. Our analysis of the ratio $f$
may be regarded as a first step to explore an index which gives the
criterion of the existence of AdS dual. Thus an interesting future
problem is to compute $f_{max}$ for all toric diagrams, and also to
see if the situation does not change when we extend the examples to
non-toric ones.

We would also like to stress that the ratio $f$ is mathematically
well-defined for all toric Sasaki-Einstein manifolds in that it
takes a unique value corresponding to a toric diagram. Thus, even
the fact that the ratio $f$ is always finite for any toric
Sasaki-Einstein manifolds is already an non-trivial mathematical
result.

Finally we would like to point out that we can define other ratios
which are computable from the central charges and field contents of
$\mathcal{N}=1$ SCFTs. One such example is the ratio
$g=\f{N_{gauge}}{4a}$ using the number $N_{gauge}$ of vector
multiplets, instead of our previous ratio $f=\f{N_{B}}{16a}$. We can
again show $g=1$ for orbifold quiver gauge theories. For example,
 in the $\mathcal{N}=1$ SCFTs dual to
$Y^{p,q}$ this new ratio $g$ takes the values within the range
$1\leq g\leq \f{32}{27}$, where the maximum is taken when $X_5$ is
(an orbifold of) $T^{1,1}$. We generally expect the range $1\leq
g\leq g_{max}$ $(g_{max}\sim O(1))$ for all $\mathcal{N}=1$ SCFTs
dual to toric Sasaki-Einstein manifolds.

Two more interesting quantities will be $h_a=\f{a_{free}}{a_{SCFT}}$
and $h_c=\f{c_{free}}{c_{SCFT}}$, which are the ratios of the
central charges $a$ and $c$ of the free Yang-Mills to the ones for
the interacting SCFT. It is trivial to see that $h_a=h_c=1$ for any
orbifold quiver gauge theories. For $Y^{p,q}$ we find the values
$1\leq h_a\leq 1.10352$ and $\f{80}{81}\leq h_c\leq 1.05321$. Notice
that $h_a$ is always greater than $1$ while $h_c$ is not in this
example. This suggests that there exist RG-flows\footnote{We would
like to thank Igor Klebanov very much for pointing out this
possibility to us.} from orbifold theories to the interacting SCFTs
since the central charge $a$ always decreases under the RG-flow,
while $c$ does not.

\vskip2mm

\noindent {\bf Acknowledgments}

TT would like to thank H. Isono, I. Klebanov, T. Oota, K. Skenderis,
Y. Tachikawa,
 M. Yamazaki and Y. Yasui for providing us important comments and
 references.
TN is grateful to the members of Theoretical Particle Physics Group in
Kyoto Univ. for useful discussions and encouragements during the
preparations of his master thesis, part of which is based on the
present paper. The work of TT is supported in part by JSPS
Grant-in-Aid for Scientific Research No.18840027.

\vskip2mm




\begin{thebibliography}{99}

\baselineskip=8pt


\bibitem{Maldacena}
 J.~M.~Maldacena,
  ``The large N limit of superconformal field theories and supergravity,''
  Adv.\ Theor.\ Math.\ Phys.\  {\bf 2}, 231 (1998)
  [Int.\ J.\ Theor.\ Phys.\  {\bf 38}, 1113 (1999)]
  [arXiv:hep-th/9711200];
  O.~Aharony, S.~S.~Gubser, J.~M.~Maldacena, H.~Ooguri and Y.~Oz,
  ``Large N field theories, string theory and gravity,''
  Phys.\ Rept.\  {\bf 323}, 183 (2000)
  [arXiv:hep-th/9905111].

\bibitem{HeSk}
  M.~Henningson and K.~Skenderis,
  ``The holographic Weyl anomaly,''
  JHEP {\bf 9807} (1998) 023
  [arXiv:hep-th/9806087].

\bibitem{GMSW}
  J.~P.~Gauntlett, D.~Martelli, J.~Sparks and D.~Waldram,
  ``Sasaki-Einstein metrics on S(2) $\times$ S(3),''
  Adv.\ Theor.\ Math.\ Phys.\  {\bf 8}, 711 (2004)
  [arXiv:hep-th/0403002].

\bibitem{BFHMS}
  S.~Benvenuti, S.~Franco, A.~Hanany, D.~Martelli and J.~Sparks,
  ``An infinite family of superconformal quiver gauge theories with
  Sasaki-Einstein duals,''
  JHEP {\bf 0506}, 064 (2005)
  [arXiv:hep-th/0411264].


\bibitem{CLPP}  M.~Cvetic, H.~Lu, D.~N.~Page and C.~N.~Pope,
 ``New Einstein-Sasaki spaces in five and higher dimensions,''
  Phys.\ Rev.\ Lett.\  {\bf 95}, 071101 (2005)  [arXiv:hep-th/0504225].


\bibitem{KlWi}
  I.~R.~Klebanov and E.~Witten,
  ``Superconformal field theory on
   threebranes at a Calabi-Yau  singularity,''
  Nucl.\ Phys.\  B {\bf 536}, 199 (1998)
  [arXiv:hep-th/9807080].

\bibitem{GG}
  S.~Benvenuti, B.~Feng, A.~Hanany and Y.~H.~He,
  ``Counting BPS operators in gauge theories: Quivers, syzygies and
  plethystics,''
  arXiv:hep-th/0608050;
  D.~Martelli and J.~Sparks,
  ``Dual giant gravitons in Sasaki-Einstein backgrounds,''
  Nucl.\ Phys.\  B {\bf 759} (2006) 292
  [arXiv:hep-th/0608060];
  A.~Basu and G.~Mandal,
  ``Dual giant gravitons in AdS(m) x Y**n (Sasaki-Einstein),''
  arXiv:hep-th/0608093.


\bibitem{GuKlPe}
S.~S.~Gubser, I.~R.~Klebanov and A.~W.~Peet,
  ``Entropy and Temperature of Black 3-Branes,''
  Phys.\ Rev.\  D {\bf 54}, 3915 (1996)
  [arXiv:hep-th/9602135].

\bibitem{Wi}
  E.~Witten,
  ``Anti-de Sitter space, thermal phase transition, and confinement in  gauge
  theories,''
  Adv.\ Theor.\ Math.\ Phys.\  {\bf 2} (1998) 505
  [arXiv:hep-th/9803131].





\bibitem{Hage}
  O.~Aharony, J.~Marsano, S.~Minwalla, K.~Papadodimas and M.~Van Raamsdonk,
   ``The Hagedorn / deconfinement phase transition in weakly coupled large N
  gauge theories,''
  Adv.\ Theor.\ Math.\ Phys.\  {\bf 8}, 603 (2004)
  [arXiv:hep-th/0310285];
  B.~Sundborg,
  ``The Hagedorn transition, deconfinement and N = 4 SYM theory,''
  Nucl.\ Phys.\ B {\bf 573}, 349 (2000)
  [arXiv:hep-th/9908001].






\bibitem{MaSpYa}
  D.~Martelli, J.~Sparks and S.~T.~Yau,
  ``The geometric dual of a-maximisation for toric Sasaki-Einstein
  manifolds,''
  Commun.\ Math.\ Phys.\  {\bf 268} (2006) 39
  [arXiv:hep-th/0503183];
  D.~Martelli, J.~Sparks and S.~T.~Yau,
  ``Sasaki-Einstein manifolds and volume minimisation,''
  arXiv:hep-th/0603021.

\bibitem{IW}
  K.~Intriligator and B.~Wecht,
  ``The exact superconformal R-symmetry maximizes a,''
  Nucl.\ Phys.\  B {\bf 667} (2003) 183
  [arXiv:hep-th/0304128].

  \bibitem{BZ}
  A.~Butti and A.~Zaffaroni,
  ``R-charges from toric diagrams and the equivalence of a-maximization and
  Z-minimization,''
  JHEP {\bf 0511} (2005) 019
  [arXiv:hep-th/0506232];
``From toric geometry to quiver gauge theory: The equivalence of
  a-maximization and Z-minimization,''
  Fortsch.\ Phys.\  {\bf 54} (2006) 309
  [arXiv:hep-th/0512240].


\bibitem{Tac}
  Y.~Tachikawa,
  ``Five-dimensional supergravity dual of a-maximization,''
  Nucl.\ Phys.\  B {\bf 733} (2006) 188
  [arXiv:hep-th/0507057].

\bibitem{BGIW}
  E.~Barnes, E.~Gorbatov, K.~Intriligator and J.~Wright,
  ``Current correlators and AdS/CFT geometry,''
  Nucl.\ Phys.\  B {\bf 732} (2006) 89
  [arXiv:hep-th/0507146].

\bibitem{FHKVW}  S.~Franco, A.~Hanany, K.~D.~Kennaway,
D.~Vegh and B.~Wecht,  ``Brane dimers and quiver gauge theories,''
  JHEP {\bf 0601}, 096 (2006)  [arXiv:hep-th/0504110].

\bibitem{FHMSVW}  S.~Franco, A.~Hanany, D.~Martelli,
J.~Sparks, D.~Vegh and B.~Wecht,
  ``Gauge theories from toric geometry and brane tilings,''
    JHEP {\bf 0601}, 128 (2006)  [arXiv:hep-th/0505211].

\bibitem{HV}
  A.~Hanany and D.~Vegh,
  ``Quivers, tilings, branes and rhombi,''
  arXiv:hep-th/0511063.


\bibitem{FHKV}
  B.~Feng, Y.~H.~He, K.~D.~Kennaway and C.~Vafa,
  ``Dimer models from mirror symmetry and quivering amoebae,''
  arXiv:hep-th/0511287.


\bibitem{Im}
  Y.~Imamura,
  ``Anomaly Cancellations In Brane Tilings,''
  JHEP {\bf 0606} (2006) 011;
  ``Global symmetries and 't Hooft anomalies in brane tilings,''
  JHEP {\bf 0612}, 041 (2006)
  [arXiv:hep-th/0609163].

\bibitem{IIKY}
  Y.~Imamura, H.~Isono, K.~Kimura and M.~Yamazaki,
  ``Exactly marginal deformations of quiver gauge
   theories as seen from brane
  tilings,''
  arXiv:hep-th/0702049.




\bibitem{Gu}
S.~S.~Gubser,
  ``Einstein manifolds and conformal field theories,''
  Phys.\ Rev.\  D {\bf 59}, 025006 (1999)
  [arXiv:hep-th/9807164].

\bibitem{Cardy}
  J.~L.~Cardy,
  ``Is There a c Theorem in Four-Dimensions?,''
  Phys.\ Lett.\  B {\bf 215} (1988) 749.

\bibitem{BHK}
  S.~Benvenuti, A.~Hanany and P.~Kazakopoulos,
  ``The toric phases of the Y(p,q) quivers,''
  JHEP {\bf 0507} (2005) 021
  [arXiv:hep-th/0412279].

\bibitem{HoMy}
G.~T.~Horowitz and R.~C.~Myers,
  ``The AdS/CFT correspondence and a new positive energy conjecture for
  general relativity,''
  Phys.\ Rev.\  D {\bf 59}, 026005 (1999)
  [arXiv:hep-th/9808079].


\bibitem{RyTaL}
  S.~Ryu and T.~Takayanagi,
  ``Aspects of holographic entanglement entropy,''
  JHEP {\bf 0608} (2006) 045
  [arXiv:hep-th/0605073];
  S.~Ryu and T.~Takayanagi,
  ``Holographic derivation of entanglement entropy from AdS/CFT,''
  Phys.\ Rev.\ Lett.\  {\bf 96} (2006) 181602
  [arXiv:hep-th/0603001].


\bibitem{NiTa}
  T.~Nishioka and T.~Takayanagi,
  ``AdS bubbles, entropy and closed string tachyons,''
  JHEP {\bf 0701} (2007) 090
  [arXiv:hep-th/0611035].

\bibitem{Ka}
  A.~Kato,
  ``Zonotopes and four-dimensional superconformal field theories,''
  arXiv:hep-th/0610266.

\bibitem{FuOn}
  A.~Futaki, H.~Ono and G.~Wang,
  ``Transverse Kahler geometry of Sasaki manifolds and toric Sasaki-Einstein
  manifolds,''
  arXiv:math.dg/0607586;
K.~Cho, A.~Futaki and H.~Ono, ''Uniqueness and Examples of Compact
Toric Sasaki-Einstein Metrics,'' arXiv:math.dg/0701122.



 \bibitem{PLpqr}
  S.~Benvenuti and M.~Kruczenski,
  ``From Sasaki-Einstein spaces to quivers via BPS geodesics: $L(p,q|r)$,''
  JHEP {\bf 0604} (2006) 033
  [arXiv:hep-th/0505206];
  A.~Butti, D.~Forcella and A.~Zaffaroni,
  ``The dual superconformal theory for $L(p,q,r)$ manifolds,''
  JHEP {\bf 0509} (2005) 018
  [arXiv:hep-th/0505220].

\bibitem{HKW}
A.~Hanany, P.~Kazakopoulos and B.~Wecht, ``A new infinite class of
quiver gauge theories,'' JHEP {\bf 0508} (2005) 054
[arXiv:hep-th/0503177].



\bibitem{OoYa}
  T.~Oota and Y.~Yasui,
  ``New example of infinite family of quiver gauge theories,''
  Nucl.\ Phys.\  B {\bf 762}, 377 (2007)
  [arXiv:hep-th/0610092].

\bibitem{BuFoZa}
 A.~Butti, D.~Forcella and A.~Zaffaroni,
  ``Deformations of conformal theories and non-toric quiver gauge theories,''
  arXiv:hep-th/0607147.

 \bibitem{AFGJ}
  D.~Anselmi, D.~Z.~Freedman, M.~T.~Grisaru and A.~A.~Johansen,
  ``Nonperturbative formulas for central functions of supersymmetric gauge
  theories,''
  Nucl.\ Phys.\  B {\bf 526} (1998) 543
  [arXiv:hep-th/9708042].

\bibitem{AEFJ}
  D.~Anselmi, J.~Erlich, D.~Z.~Freedman and A.~A.~Johansen,
  ``Positivity constraints on anomalies in supersymmetric gauge theories,''
  Phys.\ Rev.\  D {\bf 57} (1998) 7570
  [arXiv:hep-th/9711035].







\end{thebibliography}
\end{document}